\documentclass[12pt]{article}

\usepackage[utf8]{inputenc}
\usepackage[T1]{fontenc}
\usepackage[english]{babel}
\usepackage{amsmath}
\usepackage{graphicx}
\usepackage{fancyhdr}
\usepackage{hyperref}
\usepackage{multirow}
\usepackage{authblk}
\usepackage{rsc}
\usepackage{float}
\usepackage[margin=1.2in]{geometry}
\title{Artificial intelligence real-time prediction and physical interpretation of atomic binding energies in nano-scale metal clusters}
\author[1,2]{Philomena Schlexer Lamoureux}
\author[1,2]{Tej S. Choksi}
\author[1,2]{Verena Streibel}
\author[1]{Frank Abild-Pedersen\thanks{abild@stanford.edu}}
\affil[1]{SUNCAT Center for Interface Science and Catalysis, SLAC National Accelerator Laboratory, 2575 Sand Hill Road, Menlo Park, CA 94025, United States}
\affil[2]{Department of Chemical Engineering, Stanford University, 443 Via Ortega, Stanford, CA 94305, United States}

\date{April 2020}
\begin{document}
\maketitle

\begin{abstract}
Single atomic sites often determine the functionality and performance of materials, such as catalysts, semi-conductors or enzymes. Computing and understanding the properties of such sites is therefore a crucial component of the rational materials design process. Because of complex electronic effects at the atomic level, atomic site properties are conventionally derived from computationally expensive first-principle calculations, as this level of theory is required to achieve relevant accuracy. In this study, we present a widely applicable machine learning (ML) approach to compute atomic site properties with high accuracy in real time. The approach works well for complex non-crystalline atomic structures and therefore opens up the possibility for high-throughput screenings of nano-materials, amorphous systems and materials interfaces. Our approach includes a robust featurization scheme to transform atomic structures into features which can be used by common machine learning models. Performing a genetic algorithm (GA) based feature selection, we show how to establish an intuitive physical interpretation of the structure-property relations implied by the ML models. With this approach, we compute atomic site stabilities of metal nanoparticles ranging from 3-55 atoms with mean absolute errors in the range of 0.11-0.14 eV in real time. We also establish the chemical identity of the site as most important factor in determining atomic site stabilities, followed by structural features like bond distances and angles. Both, the featurization and GA feature selection functionality are published in open-source python modules. With this method, we enable the efficient rational design of highly specialized real-world nano-catalysts through data-driven materials screening. 
\end{abstract}

\section*{New Concepts}
We present a new way to transform atomic structures into relevant features which can be used to train conventional machine learning models. The models are used to predict atomic site stabilities, but are not inherently limited to those, i.e. any other property can be in principle computed this way. The featurization scheme captures relevant structural information that defines non-crystalline structures and therefore opens up the possibility to perform ML-based high-throughput screening on systems with finite size effects and/or complex atomic arrangements as they arise in nano-materials and materials interfaces. Hereto, ML-assisted computational high throughput screening was restricted to crystalline solids or small molecular systems, whereas intermediate nano-scale materials were inaccessible due to their inherent geometrical and electronic complexity. The featurization scheme presented is based on structural and basic chemical information only and does therefore not rely on computationally expensive first-principle simulations for predictions. In addition, we developed a method to achieve physical interpretability of any machine learning method used in combination with the featurization scheme. This method is based on a genetic algorithm based feature selection. We achieve high accuracy atomic site stability predictions with mean absolute errors of only 0.11-0.14 eV compared to first principles. We furthermore establish that the chemical identity is the most important factor driving the magnitude of the site stability, closely followed by structural factors. With increasing system size, structural factors lose importance as the atomic environment becomes more and more crystalline. The developed approach is provided as publicly available open-source software.

\section*{Keywords}
Machine learning, nano materials, scaling relations, atomic site stabilities, density functional theory

\section*{GTOC}

\begin{figure}[H]
 \centering
 \includegraphics[width=0.8\linewidth]{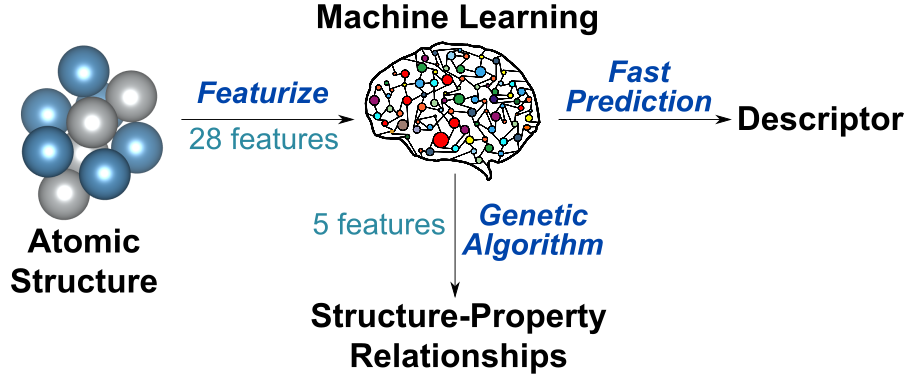}
\end{figure}

\section{Introduction} \label{introdution}
The working principle of a vast variety of material classes is governed by atomic-scale features, such as low-coordinated sites, point defects and other complex atomic arrangements on the sub-nanometer scale. Prominent examples of such materials whose functionality is governed on the atomic-scale are heterogeneous catalysts\cite{norskov2008nature, norskov2002universality, norskov2014fundamental, ma2006organic, thomas2014principles}, molecular catalysts\cite{janet2017predicting, nandy2019machine}, semiconductors\cite{chen2010intrinsic}, molecular electronics\cite{joachim2005molecular}, bio-engineered materials\cite{kasemo2002biological}, and enzymes\cite{toscano2007minimalist}. To date, meaningful computational studies of such materials require complex atomic-scale models combined with high-accuracy first-principle calculations.\cite{vojvodic2015new, norskov2002universality, norskov2006density, janet2017predicting} The knowledge generated by such studies paves the way for the rational design of high-performance materials with controlled properties. 

So, what is the implication of computing the stability of individual atomic sites in real-time? In heterogeneous catalysis, the answer to this question lies in the power of scaling relations, such as those based on adsorption energies\cite{abild2007scaling, plessow2015examining, norskov2006density, norskov2002universality, norskov2014fundamental, calle2015introducing, wang2011universal} and those based on Brønsted-Evans-Polanyi\cite{bronsted1924katalytische, evans1936further} relations. Just as the adsorption energies of different molecules on the same catalyst material linearly correlate with each other, the site stability of the adsorption site itself has been shown correlate with molecule binding energies by theory\cite{roling2018structure,choksi2019predicting, nunez2019optimization, dean2019unfolding, mpourmpakis2010identification} and experiment.\cite{campbell2013energetics, campbell2013anchored} Thus, knowledge about the site stability allows us to derive estimates on reaction kinetics and to intuitively understand the characteristics of a good catalyst and therewith develop improved catalysts.\cite{yu2016bond, schumann2018selectivity, yu2018generic, yoo2018theoretical, andersen2019beyond}. The use of site stabilities as descriptors opens therefore the possibility to screen large sets of materials for their suitability as catalysts. In this study, we specifically explore the regime of sub-nanometer and nanometer scale metallic materials, however the findings and methods hold wider potential for molecular systems\cite{janet2017predicting, nandy2019machine} and bulk materials, too.

In the last decade, machine-learning (\textbf{ML})aided computational materials screening strategies have evolved as a helpful approach to accelerate the materials discovery process.\cite{tran2018active, ma2015machine, greeley2011computational, chakraborty2017rational, studt2014discovery} However, so far, high-throughput computational screening has been applied mainly to perfect periodic bulk materials.\cite{ward2017including,montoya2017high,boes2019graph, studt2014discovery, tran2018active, janet2018resolving, curtarolo2013high, ma2015machine} In order to utilize machine learning to predict material properties, the atomic structure information has to be transformed into a format that is digestible by a ML algorithm. This is called the \textit{featurization} process. While there are some featurization methods available for perfect bulk materials, such as Voronoi tessellation\cite{ward2017including, back2019convolutional}. These cannot be applied to amorphous structures and other complex or low-coordination atomic arrangements.

Because atomic sites so often significantly define materials functionality\cite{yang2016intrinsic}, understanding and ultimately controlling their physical and chemical state holds immense potential. For this reason, a variety of site-specific descriptors have been developed, which can be categorized into (1) \textit{electronic} and (2) \textit{geometric} features. Prominent examples for electronic structure features are the d-band center\cite{dahl1999role,hammer1996co} and moments of the d-band distribution\cite{xin2014effects}, which can be computed via density functional theory (\textbf{DFT}) or the tight-binding model.\cite{bartok2013representing} The computation of electronic structure features requires the careful selection of the underlying electronic structure method (e.g. tight-binding vs. DFT vs. DFT + Hubbard U\cite{anisimov1991band}) and once established, the resulting featurization schemes have limited computational efficiency and limited generalization potential in materials space.

In terms of geometric features, notable featurization schemes include the coordination-based alloy stability model\cite{roling2017configurational,roling2019coordination}, generalized coordination numbers (\textbf{GCN})\cite{calle2015finding, calle2014fast}, orbital-wise coordination numbers\cite{ma2017orbitalwise}, and the smooth overlap atomic position (\textbf{SOAP}) approach\cite{bartok2013representing}. In the coordination-based alloy stability model, a unique $\alpha$-parameter is computed from DFT calculations for all possible coordination numbers of site, respectively. Once the parameters are known, a linear (additive) scheme is applied, based on neighbor composition and coordination number of the site. With this simple linear scheme, mean absolute errors (MAE) down to 0.03 eV can be achieved.\cite{roling2019coordination} Although corrections for compression/expansion are under development, this scheme relies mainly on coordination numbers. 

The coordination-based alloy stability model as well as the GCN and orbitalwise CN approaches work best for highly ordered systems, such as extended crystalline surfaces and nanoparticles larger than ~2 nm. Since these models have been derived from extended bulk and surface structures, they may not achieve the same accuracy for disordered structures and sub-nanometer particles, which are affected by finite- and quantum-size effects.\cite{volokitin1996quantum,li2012investigation,yudanov2002metal} Such effects can for instance result in the discretization of electronic states, giving rise to the well-known odd-even effects\cite{volokitin1996quantum, bouwen1999stability}.

The SOAP approach compares materials sections using a similarity kernel.\cite{bartok2013representing} Jinnouchi et al.\cite{jinnouchi2017predicting, jinnouchi2017extrapolating} have successfully used this approach to predict adsorption energies on nano- and sub-nanometer particles. The authors found, however, that they needed to include small nanoparticle data in their training set and could not only use slab calculations of extended surfaces to accurately predict nanoparticle energetics. Still, the accuracy for sub-nanometer particles remained limited with MAEs between 0.2-0.4 eV. For such kernel methods, the prediction accuracy systematically improves by increasing the number of training data to cover all possible local structures. Kernel methods, however, scale unfavorably with an increasing number of training instances and thus the SOAP approach will become computationally demanding when better accuracy is required. 

Non-linear machine learning (\textbf{ML}) models have been successfully used to predict chemical adsorption energies based on geometric and electronic structure features.\cite{zhao2019trends, ma2015machine, gasper2017adsorption, li2017feature, jager2018machine,ulissi2017machine, li2017high, ulissi2017address} For instance, Alexandrova et al. developed a featurization scheme based on electronic structure features.\cite{alexandrova2005search, zhai2016ensemble, alexandrova2010h, sumpter1992potential} As mentioned before, however, the use of electronic structure features is computationally demanding and defeats the purpose of a fast predictive artificial intelligence. For an in-depth overview on featurization schemes and machine learning techniques in computational heterogeneous catalysis, the reader is referred to our recent review on this topic.\cite{schlexer2019machine}
 
In short, to date site-specific studies either use relatively simple linear coordination-based models, which work best for crystalline metallic structures and fail for smaller nano-particles and clusters or they are based on non-linear approaches which use electronic structure features, which are expensive to compute. The challenge is to use "cheap" features to compute site-specific properties of sub-nanometer, non-crystalline or fluxional structures, such as they arise in relevant materials under realistic operating conditions.

In this study, we present a site-specific featurization scheme that transforms atomic structures into features, which we use in combination with machine learning (\textbf{ML}) algorithms to predict atomic site-stabilities with high accuracy (mean absolute errors as low as 0.1 eV). The featurization scheme is applicable to non-periodic and periodic structures as well as non-ordered (amorphous) and ordered (crystalline) structures. The scheme relies only on basic chemical and geometric information, hence it circumvents the need for expensive electronic structure calculations. Furthermore, this scheme is equally applicable to diverse materials (metals, oxides, molecules) as well as structural classes (surfaces, nano-materials, molecules, interfaces). Specifically, the featurization algorithm takes the atomic structure and the site index of the site of interest (e.g. active site candidate) as inputs. Based on this, the algorithm computes a set of features that represents the local chemical environment of the site of interest as well as a set of non-local, system-wide features. The generated features are well suited for conventional linear and non-linear ML models and therewith achieve high-accuracy predictions. We further present a genetic algorithm for feature selection to achieve physical interpretability of the ML algorithms, revealing structure-property relationships. All data handling, featurization, and feature selection functionalities used in this work are provided to the community in free open-source python modules. We successfully apply the developed functionality to predict atomic site stabilities of atoms in sub-nanometer and nanometer particles, achieving MAEs of as low as 0.1 eV using only 5 features based on geometry and basic chemical information. Finally, we compare the approach to a previously published scheme ($\alpha$-scheme) and discuss the pros and cons of the two approaches for different system sizes.

\section{Computational Details}\label{methodology}
We created and optimized atomic structures of metal sub-nanometer particles of 3-13 atoms, cuboctahedral nanoparticles of 55 atoms and surfaces using first principles as described in detail in the supporting information. The featurization scheme and the genetic algorithm feature selection, are explained in detail as well. The source code for the featurizer and the GA feature selection is open-source available at:
\hyperlink{github.com/schlexer/CatLearn}{github.com/schlexer/CatLearn}.

\section{Results} \label{results}
We will show how to featurize atomic sites in 3-13 atom clusters and use machine learning to predict atomic sites stabilities in section \ref{clusters}. Then, we  present the genetic algorithm for feature selection, which enables us to understand structure-property relations of atomic site in more detail and inspect the importance of different features in depth. In section
\ref{nanoparticles}, we apply the scheme to 55-atom nanoparticles. In section \ref{surfaces} we discuss the perks and drawbacks of the Ml-based approach versus a coordination-based approach for extended surfaces.

\subsection{3-13 atom clusters}\label{clusters}
\subsubsection{Featurization of atomic sites}
We generated mono- and bimetallic sub-nanometer clusters as described in the computational details. All possible unique combinations of $\mathrm{A_{x}B_{y}}$ with A, B $\mathrm{\in}$ \{Ni, Cu, Pd, Ag, Pt, Au\} and (x+y) $\mathrm{\in}$ \{3, 4, 5 \dots 13\} were computed. Briefly, atomic positions were pre-optimized using the EMT potential in combination with a genetic algorithm. The best candidates were then further relaxed using DFT. This relaxation yields stable local minima, but does not ensure global minima. In fact, we wish that our modeling strategy and derived insights are not restricted to global minimum structures, given the dynamic nature (fluxionality\cite{zhai2017fluxionality, sun2019theoretically}) of working catalysts and the catalytic importance of local minima structures like e.g. stepped surfaces. 

We randomly selected atomic sites from the set of all possible atoms from all monometallic and bimetallic sub-nanometer particles. In order to predict the stability of these atomic sites, we analyze their features, i.e. properties. We can conceptually distinguish between \textit{system-specific} features (e.g. stoichiometry of the nanoparticle, particle size (\mbox{\#} atoms)), and \textit{site-specific} features (e.g. the coordination number or chemical environment of the site). Furthermore, we can distinguish between \textit{physical/chemical} and \textit{structural} features. Physical/chemical features are for instance atomic numbers, valence electrons, or electro-negativity, whereas structural features entail metrics of interatomic distances, angles, or coordination numbers. 

\begin{figure}[H]
 \centering
 \includegraphics[width=0.8\linewidth]{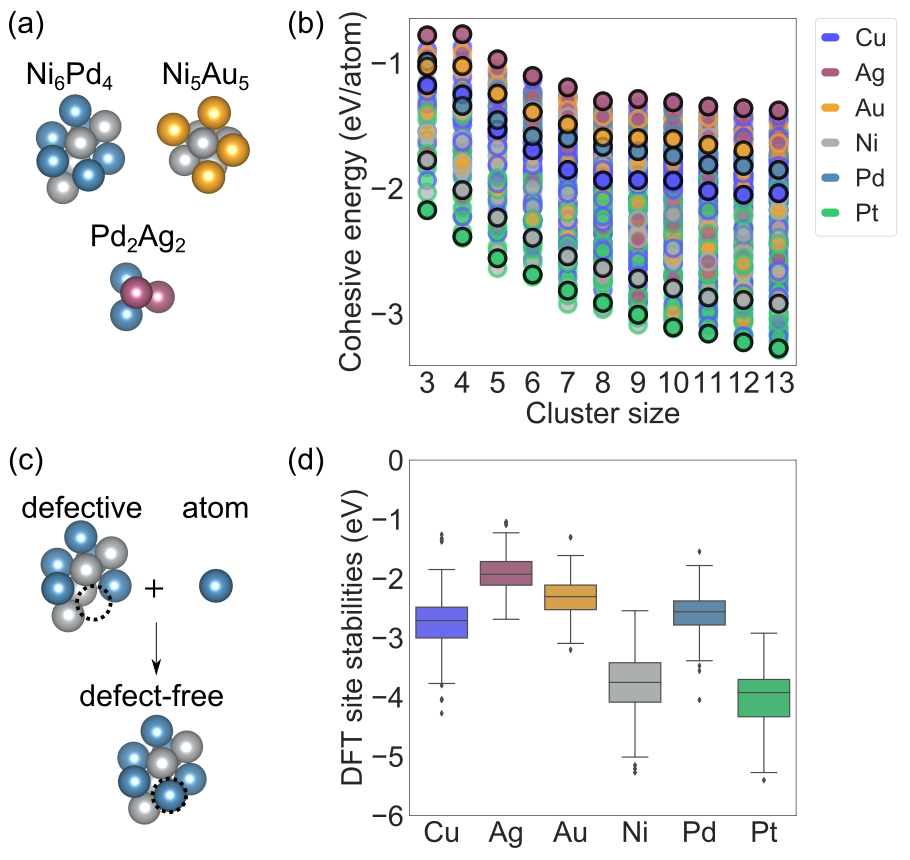}
 \caption{Cohesive energies and site stabilities of 3-13 atom clusters. (a) A selection of optimized clusters. (b) Cohesive energies of all defect-free clusters. For example, a green dot surrounded by  a red circle denotes a Pt-Ag bimetallic cluster. The cluster size is x+y, where x (y) is the number of atoms of element A (B), and black circles denote monometallic clusters. (c) Site stability computation scheme: Negative energies denote exothermic site stabilities. (d) Boxplots of site stabilities by element of site. }
 \label{fig1}
\end{figure}

Our goal is to predict site stabilities from features that are determined in a computationally efficient way. We chose a set of features that consists only of structural (distances, angles, coordination numbers, etc.) and basic chemical information (atomic number, valence electrons, stoichiometric coefficients, etc.). This choice results in \textbf{28 unique features}. We abstain from using the EMT site stability as a feature since the EMT potentials are available for only a limited set of elements. The features are summarized in Fig. SI-5, and Table 1.
\newline
\newline
\begin{tabular}{p{2.2cm}|p{9cm}|p{0.5cm}|p{0.5cm}}
\hline
 \multicolumn{4}{p{13.2cm}}{Table 1: Features of $\mathrm{A_{N_1}B_{N_2}}$ binary systems. \textbf{n} denotes the number of features in this category, \textbf{t} denotes the category type: \textbf{s} denotes site-specific, \textbf{g} denotes \textit{"global"}, i.e. system-specific.} \\
 \hline
 Feature(s)    &  Description & n & t\\
 \hline
 $\mathrm{N_{tot}}$ & Total number of atoms in the system & 1 & g\\
 $\mathrm{N_{1}}$, $\mathrm{N_{2}}$   & Stoichiometric coefficients of A or B & 2 & g\\
 $\mathrm{Z_{1}}$, $\mathrm{Z_{2}}$    & Atomic number of A or B & 2 & g\\
 $\mathrm{Z_{mean}}$      & Mean of atomic numbers of all atoms  & 1 & g\\
 $\mathrm{Val_{1}}$, $\mathrm{Val_{2}}$     & Valence electrons of A or B & 2 & g\\
 $\mathrm{Val_{mean}}$      & Mean of valence electrons of all atoms  & 1 &g\\
 \hline
 $\mathrm{Z_{site}}$   & Site atomic number & 1 & s\\
 $\mathrm{Val_{site}}$   & Site valence electrons & 1 & s\\
 $\mathrm{CN_{site}}$    & Site coordination number & 1& s\\
 \hline
 $\mathrm{Z_{x, neigh.}}$  &  Metrics x of neighbor atomic numbers* & 4& s\\
 $\mathrm{CN_{x, neigh.}}$    & Metrics x of neighbor coordination numbers* & 4& s\\
 $\mathrm{\gamma_{x}}$     & Metrics x of the angles between the atomic site and each unique pair of two neighbors* & 4& s\\
 $\mathrm{d_{x}}$     & Metrics x of the distances between the atomic site and each neighbor* & 4& s\\
 \hline
 \multicolumn{4}{c}{*x can be mean, min, max or the standard deviation ($\mathrm{\sigma}$)} \\
\end{tabular}

\subsubsection{Model selection: Algorithms and features}
Having introduced our featurization schemes, we now investigate the performance of various machine learning models to predict site stabilities. The models include ordinary linear regression (also referred to as ordinary least squares, \textbf{OLS}), Gaussian process regression\cite{rasmussen2006cki} (\textbf{GPR}), neural networks (\textbf{NN}), random forests (\textbf{RF}), and extreme gradient boost\cite{chen2016xgboost} (\textbf{XGB}) decision trees, see supplementary information for more details. In order to make sure that our training set is sufficiently large for all of these algorithms, we tested the convergence of our performance metrics with training set size, see supporting information Figure SI-3. All models reach a plateau in performance after 300-400 data points. We split our data set with a test/training ratio of 240/958, and therewith achieve a sufficiently large training set. The training set is therewith large enough to produce meaningful results for our problem of interest. We optimized the model using 4-fold cross validation (\textbf{4f-cv}) on the training set, in combination with hyper-parameter optimization from python. 

\begin{figure}[H]
 \centering
 \includegraphics[width=\linewidth]{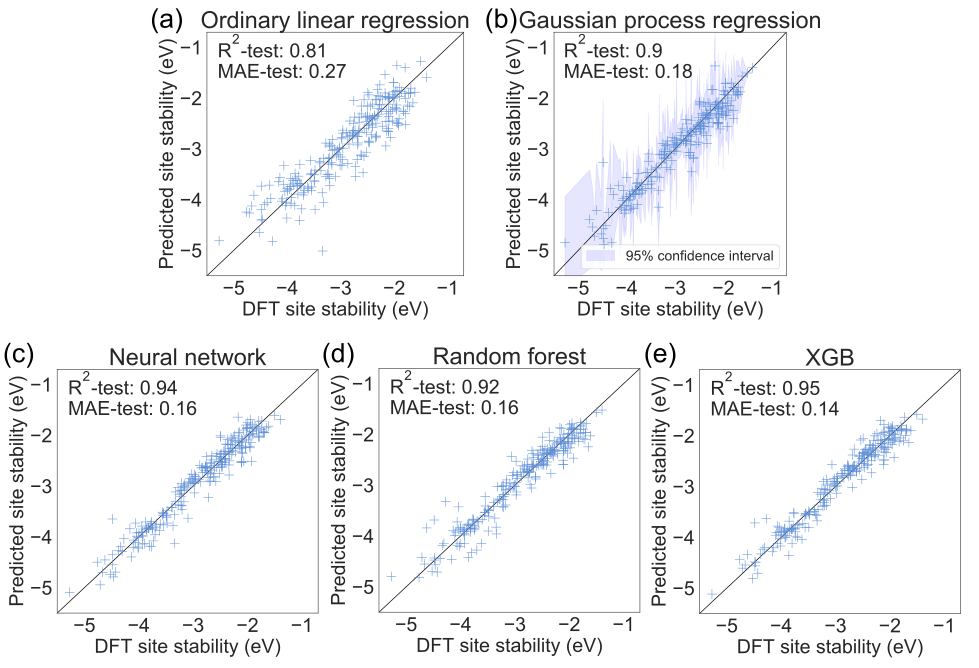}
 \caption{Performance of various models on the test set (test/training = 240/958) using all 28 features, all MAE values are given in eV. (a) Ordinary linear regression model (\textbf{OLS}). (b) Gaussian process regression (\textbf{GPR}) with 95\% confidence interval based on posterior distribution. (c) Neural network (\textbf{NN}). (d) Random forest (\textbf{RF}). (e) Extreme gradient boost (\textbf{XGB}).}
 \label{parity}
\end{figure}

Using all 28 features and fitting the most promising models on the full training set, we achieve predictions with MAEs between 0.14 and 0.27 eV. The results are shown in Figure \ref{parity}. Clearly, the neural network and the extreme gradient boost decision trees are the best performing models with an $\mathrm{R^2}$ on the test set of 0.94 and 0.95, respectively.

\subsubsection{Feature importance analysis}
\begin{figure}[H]
 \centering
 \includegraphics[width=0.7\linewidth]{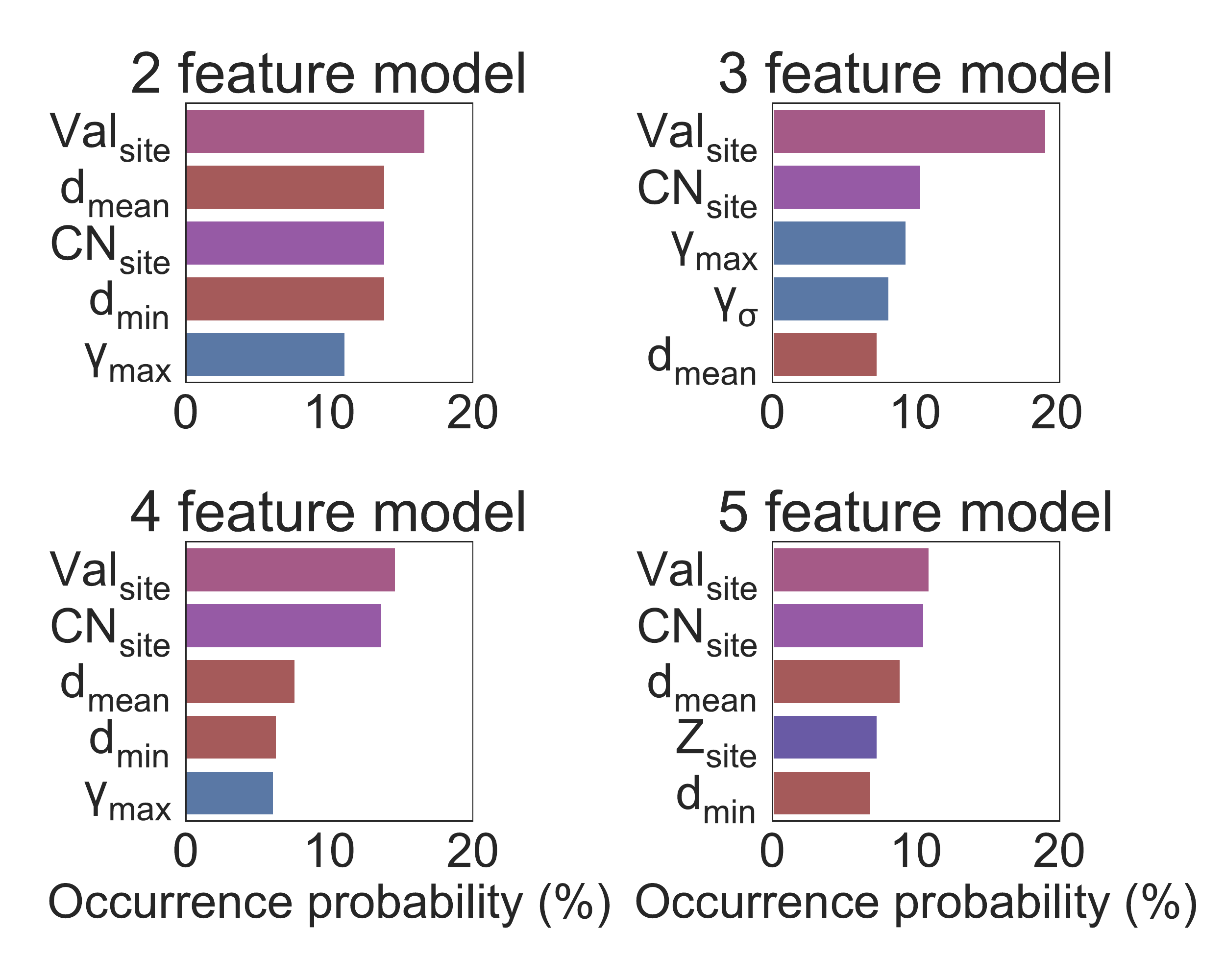}
 \caption{Systematic investigations of feature importance in models containing between 2 and 5 features. Feature occurrence probability (\%) in the set of linear models with a mean 4f-cv $\mathrm{R^2}$ in the top 5 percentile.}
 \label{orl_feature_selection}
\end{figure}

Besides evaluating different ML model, we are also interested in understanding which of the features are most important to predict the site stability. We start out the feature analysis based on linear models to inform our understanding on individually relevant features, i.e. disregarding synergies between 2 or more features. In order to analyze the importance of features, we generated a model for every possible combination of 2 features and determined each model's mean $\mathrm{R^{2}}$ in 4-fold cross-validation (\textbf{4f-cv}) on the training set. We then ranked the resulting models according to their 4f-cv. We analyzed the occurrence probability of each individual feature across all models performing in the top 5\% of 4f-cv $\mathrm{R^{2}}$. The top 5 most occurring features are shown in Figure \ref{orl_feature_selection} (upper left panel). We repeated the process for models with more features, i.e. every possible combination of p $\mathrm{\in}$ \{2,3,4,5\} out of $\mathrm{p_{tot}=28}$. The results are shown in Figure \ref{orl_feature_selection}.

Independent of the number of features in the model, the number of valence electrons of the site's atomic element in oxidation state 0, $\mathrm{Val_{site}}$, is the feature that stands out as most frequent with an occurrence probability of 10-20\% in the top 5\% of models. The second most frequent feature in all cases is the coordination number of the site, $\mathrm{CN_{site}}$. Thus, accounting for linear correlation of a feature with the target only, the number of valence electrons in combination with the coordination number show great predicting power. Other important features represent measures of distances and angles. In this feature space, the mean distance of the site to its neighbors $\mathrm{d_{mean}}$ and the maximum angle between two neighbors and the site $\mathrm{\gamma_{max}}$ stand out as important features.  

A few features are linearly correlated, such as the distance measures ($\mathrm{d_{min}}$/$\mathrm{d_{max}}$/$\mathrm{d_{\sigma}}$) and angles/coordination numbers($\mathrm{\gamma_{mean}}$/$\mathrm{\gamma_{max}}$/$\mathrm{\gamma_{\sigma}}$/$\mathrm{CN_{site}}$), see Pearson correlation matrix in the supporting information Figure SI-2. Therefore, the solutions of ordinary linear regression (and those of other models) may not be unique. These linear correlations are not problematic, though, since we do not analyze and use the best model. Rather, we quantify which features occur most often in the top 5\% of models to identify the main features determining the site stability. Since we used the linear model in this analysis, we were able to include all possible combinations of p features, as fitting the model and determining the performance metrics is computationally fast. Non-linear models and especially neural networks are computationally more expensive in training and prediction. 

We further have the goal to develop a method that is applicable to a variety of problems, including those in which not all models (feature combinations + algorithm) can be computed. We therefore developed a genetic algorithm (\textbf{GA}) for model selection. Briefly, the GA creates a population of chromosomes. A chromosome is a list of p features, which are called genes in this context. A 3-gene chromosome can for instance consist of \{$\mathrm{d_{mean}}$, $\mathrm{\gamma_{max}}$, and $\mathrm{Val_{site}}$\}, with the genes being the individual features. The GA determines the fitness of all chromosomes given a certain regressor class (e.g. XGB or linear model) and selects the best performing model via evolution over a certain number of generations. The working principle is described in more detail in the supporting information.

We chose to investigate models with 5 features (genes) at a time to reduce overall feature correlation while allowing higher-order feature interactions in non-linear models. Most feature combinations with p=5 exhibit a sum of pairwise feature correlations > 1, Figure SI-XX. As the NN and the XGB performed best when we used all features (Figure \ref{parity}), we chose to run the GA with these two models. We used a population size of 200 and an offspring size of 20. To make the neural network GA faster, we used an offspring size of 10. The larger the offspring size, the faster the algorithm evolves. 

\begin{figure}[H]
 \centering
 \includegraphics[width=0.99\linewidth]{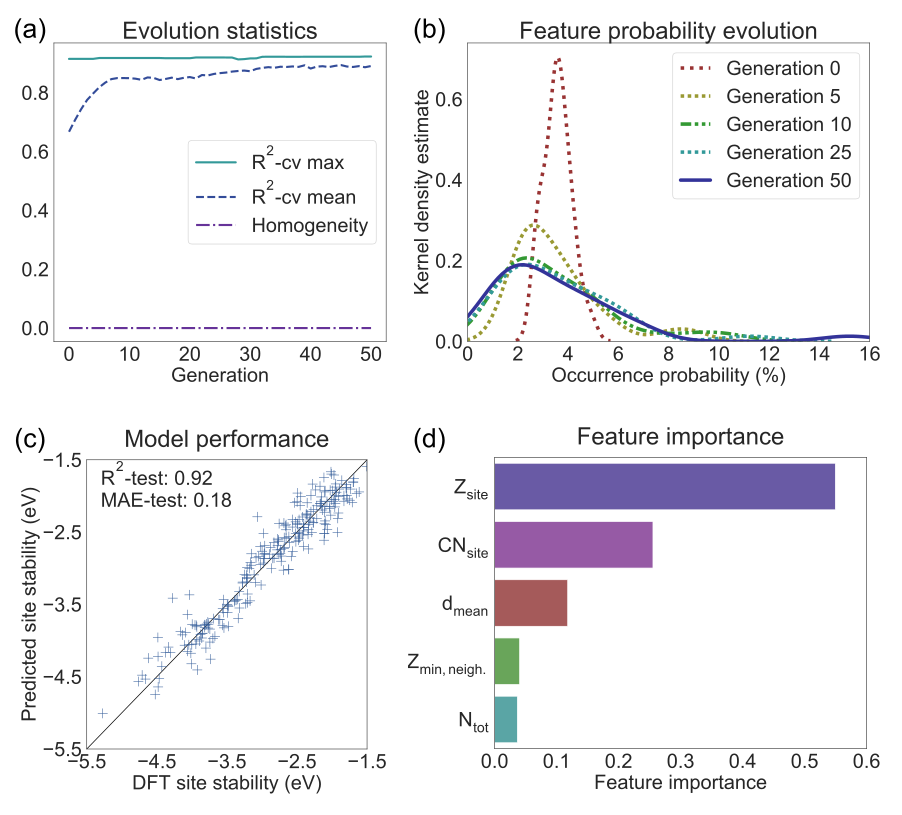}
 \caption{Genetic algorithm evolution of feature selection process using the XGB regressor. (a) Solid line: Max fitness (fitness = mean 4f-cv $\mathrm{R^2}$ of best performing chromosome). Dashed line: mean population fitness. Dot-dashed line: Homogeneity (a value of 0 indicates that all possible 28 features are available in the population). (b) Evolution of feature occurrence probability. At Generation 0, all features show an occurrence probability close to 1/28 $\approx$ 3.57\%. During evolution strong features become more probable, and weak features become less probable. (c) Best performing 5 feature model after evolution, achieving a $\mathrm{R^2}$ of 0.92 (MAE=0.18 eV) on the test set. (d) Feature importances of the 5 features used by the best model.}
 \label{ga_xgb}
\end{figure}

An in-depth analysis of the GA process using the XGB-based process is summarized in Figure \ref{ga_xgb}. In Figure \ref{ga_xgb} (a), the mean and max 4f-cv $\mathrm{R^2}$ throughout the GA evolution are shown. At generation 0, the population consists of 200 unique random chromosomes, which all show different fitness measured by the mean 4f-cv $\mathrm{R^2}$. At generation 0, the best performing chromosome (5-feature model) shows a fitness close to 0.9 (solid line) and the population shows a mean fitness of around 0.7 (dashed line). We furthermore follow the homogeneity, which indicates whether all features are still in the population. A value of 0 means that all features are available in the population. The exact definition is given in the supporting information. 

During the GA evolution, the population performance increases, resulting in an improvement of the mean population fitness in Figure \ref{ga_xgb} (a). At the same time, the occurrence frequency of features changes noticeably in Figure \ref{ga_xgb} (b). At generation 0, all features show an occurrence probability of around 1/p = 1/28 $\approx$ 3.57\%. Given the randomly created starting population of the chromosome generation, we see a tight distribution around the ideal value of 1/28. In the initial phase of the GA evolution, weak feature combinations are depleted while stronger combinations are enriched, leading to a broadening of the Gaussian distribution. In the later phase of the evolution (generations 25-50) the distribution becomes asymetric with a broader tail towards larger occurrence probabilities (6-16\%). After 50 generations, a small broad peak at around ~15\% appears. This peak represents the most relevant features. 

After an evolution of 50 generations, the best XGB model (chromosome) using 5 features (genes) achieves an $\mathrm{R^2}$ of 0.92 (MAE=0.18 eV) on the test set, see parity plot in Figure \ref{ga_xgb} (c). The 5 features used by this model are shown in Figure \ref{ga_xgb} (d), together with their feature importance. The best neural network model using 5 features achieves a $\mathrm{R^2}$ of 0.89 (MAE=0.20 eV) on the test set. The final population feature occurrence probability breakdown is shown in Figure SI-5. 

Both model types (NN and XGB) suggest a chemical feature to be most important in determining the site stability. For the NN, this is the site valence electrons ($\mathrm{Val_{site}}$) and for the XGB, the site atomic number ($\mathrm{Z_{site}}$). In both cases, the chemical measure is immediately followed by structural features like e.g. the coordination number of the site ($\mathrm{CN_{site}}$). This trend was also found using linear models (compare Figure \ref{orl_feature_selection} and Figure SI-5).

\subsection{Model extrapolation to larger nanoparticles}\label{nanoparticles}
We have seen that the machine learning models predict atomic site stabilities in sub-nanometer clusters with impressive accuracy. As open question remains, in how far these models, trained on the sub-nanometer particles data set, can extrapolate to other material types, such as larger particles. To address this question, we created 30 bimetallic, cuboctrahedral nanoparticles of stoichiometry $\mathrm{A_{x}B_{y}}$ with x=28 and y=27 and $\mathrm{A\neq B}$. For these nanoparticles, we computed the site stability of structurally distinct surface sites, systematically including corners, edges, and terraces. The resulting data set has 327 sites and the same 28 features as above. 

Seeking for large generalized models, we tried to predict nano-particle site stabilities with the XGB model trained on sub-nanometer clusters. However, we only achieve a $\mathrm{R^{2}}$ of 0.44 and a MAE of 0.47 eV using the complete nano-particle data set. This suggest that the models have to be re-trained on the nano-particle data set, which is not surprising: The influence of quantum- and finite-size effects on the 3-13 atom clusters is simply too pronounced to capture the properties of the larger, 55-atom nanoparticles. Fortunately, based on our training set convergence tests, we anticipate that training a ML model on a few hundred data points for a respective material class should be sufficient to obtain a performance of practical use. We refitted machine learning models on the 55-atom nanoparticle data set. We split the data set into training/test sets with a ratio of 261/66. As the data set is rather small, the performance metrics may show a larger variance. To quantify the variance, we computed the mean and standard deviation on the $\mathrm{R^{2}}$ in a 5-fold cross-validation on the training set. As the random forest (RF) and the extreme gradient boost (XGB) algorithm have proven fast and accurate for the sub-nanometer particles, we will continue our study with these algorithms only and abstain from performing an in-depth model comparison. 

Training the random forest (with cross-validation based hyper-parameter optimization) using all 28 features gives a $\mathrm{R^{2}_{cv}=0.95}$ (MAE=0.14 eV) on the test set. Interestingly, the 5 most important features are mainly related to the chemistry of the site and the nanoparticle stoichiometry ($\mathrm{Val_{site}}$/$\mathrm{Val_{1}}$/$\mathrm{Z_{site}}$/$\mathrm{Val_{mean}}$/$\mathrm{CN_{site}}$). We verified this outcome using different random states for the training set. 

As for the sub-nanometer particles, the XGB algorithm outperforms the RF with an $\mathrm{R^{2}}$ of 0.97 (MAE=0.11 eV) on the test set. The parity plot and the feature importances (NP, XGB, all 28 features) are shown in Figure SI-6 (a-b). In alignment with the RF model, the XGB model uses various chemical parameters as the most important features. This finding shows how small structural variance within a material class results in the chemical features being more relevant, which ties in nicely with simple coordination-based models for larger nano-particles and surfaces.\cite{choksi2019predicting} However, we note that also for the sub-nanometer particles, chemical information (like atomic number and valence electrons) has always ranked as top feature, followed by structural features. 

As the chemical features used in the best nano-particle models are slightly correlated, we anticipate that reducing the feature space could still lead to very precise results for the nanoparticles. To confirm this, we performed a model selection using the genetic algorithm introduced in sections \ref{methodology} and \ref{clusters}. After 50 generations of evolution, the best performing model using only 5 features achieves an $\mathrm{R^2}$ of 0.97 (MAE=0.11 eV) on the test set, see supporting information Figure SI-6 (c-d). That is, the 5-feature model performs just as well as the 28-feature model. 

Furthermore, the best 5-feature model for the 55-atom nano-particles has a similar feature set up as that of the clusters, with the site atomic number as most important feature, followed by structural parameters. However, the relative importance of the structural features changes, compare Figure 4 (d) and Figure SI-6 (d). In fact, the importance of structural features decreases with increasing system size. This makes sense as there is less structural variability for surface atoms in this materials class. Thus, the main distinction lies within the chemical identity of the site of interest and its neighbors. 

Overall, it is encouraging to observe such a good performance despite the relatively small data set. It also shows that the featurization scheme works for materials that are different from the sub-nanometer particles. We acknowledge that it remains to be confirmed for other material classes, such as oxides, carbides, or two-dimensional materials. The software is publicly available for the scientific community to perform such studies, see link in the computational details.

\subsection{Surfaces}\label{surfaces}
In the following, we are moving away from clusters and nanoparticles towards extended surfaces. Atoms embedded in a crystalline surface have less structural variability and, therefore, as discussed in the introduction, the coordination-based (\textbf{CN}) alloy stability model\cite{choksi2019predicting, roling2017configurational, roling2019coordination} works well to predict site stabilities of atoms in single-crystalline metal surfaces. The strength of this alloy stability model is that it works with only a few parameters and the interpretability is  more intuitive than ML-based models, although we have demonstrated in this study how ML models can yield an in-depth analysis of contributing features through (GA-based) feature selection. The main differences in the two approaches are that ML models inherently strive on increased data availability while being based on non-DFT features, whereas the CN-based models require the computation of only a few systems while using more elaborate investigation of those to manually establish the parameters.

We will now compare the coordination-based alloy stability model to the ML models introduced in this work. To this end, we computed the site stability of atoms at chemically distinct surfaces sites in (111), (211), and (100) surfaces of Ag, Au, Cu, Pd, and Pt. As we considered only monometallic systems, we removed all alloy-related features like $\mathrm{Z_{mean}}$, $\mathrm{Val_{mean}}$, $\mathrm{Z_{max, neigh.}}$, etc. from the ML models. The resulting data set consists of 70 site stabilities and 16 features, which we parted into a training/test ratio of 56/14. As this data set is very small, we compared the performance of all models (XGB, RF, NN, ORL) using all features. The tree-based models were trained using hyper-parameter optimization via 10-fold cross-validation on the training set. 

The best performing model is the XGB with a $\mathrm{R^{2}_{XGB}=0.79\pm 0.08}$ (MAE = 0.42 eV). We anticipate the performance metrics to improve significantly when the ML models are trained on larger data sets. Importantly, the CN-based alloy stability model has shown to perform well with a MAE of 0.18 eV in a similar study on alloy surfaces, while not requiring a large materials variety.\cite{roling2019coordination} This is not surprising as the alloy stability model makes use of accurate DFT-derived features tied to the site of interest.

\section{Conclusions}
In this study, we present a robust scheme to transform atomic structure data into site-specific features which can be fed into any common machine learning (ML) algorithm. Training various ML models on DFT-based site stabilities, this approach allows us to predict site stabilities of sub-nanometer and nanometer bimetallic particles with high accuracy of MAE 0.11-0.14 eV in real time.

We furthermore present a way to interpret the ML models using genetic-algorithm based feature selection. Our feature analysis shows how the importance of structural versus chemical features increases as we transition from extended surfaces to nanometer to sub-nanometer metal structures. Specifically, we establish that a chemical feature (site valence electrons ($\mathrm{Val_{site}}$ or the site atomic number ($\mathrm{Z_{site}}$)), immediately followed by structural features (such as the site coordination number ($\mathrm{CN_{site}}$), distances and angles) represents a powerful feature combination across all models and system sizes. 

Finally, we compared the ML approach to the coordination-based alloy stability model for atomic sites in crystalline metal surfaces. We find that the coordination-based alloy stability model performs well for crystalline atomic sites in surface and nanoparticles with a diameter of > 1.6 nm. While the coordination-based alloy model makes use of only a few accurate first-principle features, the ML approach thrives on data quantity and variety, which requires a larger training data set. 

Our machine learning based approach (including the featurization, feature analysis and ML models) opens the possibility to perform high-throughout screening for complex, amorphous, and sub-nanometer structures in materials which exhibit a large structural variety. 

\section*{Acknowledgements}
This work was supported by the U.S. Department of Energy, National Energy Research Scientific Computing Center (NERSC) U.S. Department of Energy Office of Science User Facility, Chemical Sciences, Geosciences, and Biosciences (CSGB) Division of the Office of Basic Energy Sciences to the SUNCAT Center for Interface Science and Catalysis.
PS and VS gratefully acknowledge the Alexander von Humboldt Foundation (AvH) for financial support.

\bibliography{main.bib}

\end{document}


\maketitle

\section*{Keywords}
Machine learning , scaling relations, atomic site stabilities, density functional theory

\section{Computational details}
\subsection{Density functional theory calculations}\label{DFT}
Periodic density functional theory (\textbf{DFT}) calculations were performed using Quantum ESPRESSO~\cite{giannozzi2009quantum}, the BEEF-vdw exchange-correlation functional\cite{wellendorff2012density} in combinations with ultra-soft Vanderbilt pseudo-potentials\cite{vanderbilt1990soft}, and the atomic simulation environment (ASE)\cite{larsen2017atomic}. Plane waves and charge densities were expanded up to a kinetic energy of 500 eV and 5000 eV, respectively. 

Mono- and bimetallic sub-nanometer clusters with 3-13 atoms were generated by randomly positioning atoms in a box of $(15\times 15\times 15)$\,\AA$^3$. The box was located in a $(30\times 30\times 30)$\,\AA$^3$ unit cell. A genetic algorithm\cite{vilhelmsen2014genetic} combined with effective medium theory (EMT) potentials was then used to search for stable particle structures. This search was performed using a starting population of 40 candidates of random atomic positions per particle stoichiometry. The lowest energy candidate in the final population of the EMT-based search was then further optimized with DFT. The DFT calculations for clusters and nano-particles were performed with a $(1\times 1\times 1)$ k-point set using the Monkhorst-Pack scheme for Brillouin-zone integration.\cite{monkhorst1976special} All possible unique combinations of $\mathrm{A_{x}B_{y}}$ with A, B $\mathrm{\in}$ \{Ni, Cu, Pd, Ag, Pt, Au\} and (x+y) $\mathrm{\in}$ \{3, 4, 5 \dots 13\} were computed.

55-atom nanoparticles were generated using the layer specification approach within ASE, considering only (111) and (100) surfaces. Specifying 2 layers in each direction resulted in a cuboctahedral, 55-atom nanoparticle. From this template, we generated monometallic nanoparticles by setting the interatomic distances to those found in the respective bulk material. For bimetallic particles, we used the lattice constant estimated via Vegard's law\cite{vegard1921konstitution} to adjust interatomic distances. 

Metal surfaces were represented by ($2\times 2$) surface unit cells of slabs consisting of 7 atomic layers for monometallic systems. For slabs, we used a ($7 \times  7\times 1$) k-point mesh and slab replicas were separated by more than 12 \AA. For the slabs and nanoparticles, all atoms were constrained to their bulk geometries.

Site stabilities of individual atoms were computed by single-point DFT calculations. The site stability of site A $\mathrm{E_{site}(A)}$ is defined in equation \eqref{stability}, where $\mathrm{E_{defect-free}}$ is the total energy of the original, defect-free cluster, $\mathrm{E_{A}}$ is the total energy of the free-standing atom A and $\mathrm{E_{defective}}$ is the original structure  without atom A:

\begin{equation}\label{stability}
    \mathrm{E_{site}(A) = E_{defect-free} - E_{A} - E_{defective}}
\end{equation}

The spin states of free-standing atoms were thereby: 
\begin{table}[H]
    \centering
    \begin{tabular}{c|c} 
    \hline
    Atom  & $\mathrm{mag/cell \ (Bohr})$ \\
    \hline
    Ag & 1.0 \\
    Au & 1.0 \\
    Cu & 1.0 \\
    Co & 3.0 \\
    Rh & 3.0 \\
    Ir & 3.0 \\
    Fe & 4.0 \\
    Os & 4.0 \\
    Mn & 5.0 \\
    Re & 5.0 \\
    Ni & 2.0 \\
    Pd & 0 \\
    Pt & 0 \\
    Ru & 0 \\
    \hline
    \end{tabular}
\caption{Spin states of atoms in vacuum (box of $(15\times 15\times 15)$\,\AA$^3$).\label{atoms}}
\end{table}

The cohesive energy $\mathrm{E_{coh}}$ of a cluster $\mathrm{A_{x}B_{y}}$ is computed as shown in equation \eqref{cohesive}, where $\mathrm{E_{defect-free}}$ is the total energy of the defect-free cluster, $\mathrm{E_{A/B}}$ is the total energy of the free-standing atom A/B, and x and y are stoichiometric coefficients. As the cohesive energy is given per atom, it can be interpreted as mean site stability of all atoms in the cluster.

\begin{equation}\label{cohesive}
    \mathrm{E_{coh}(A_{x}B_{y}) = (E_{defect-free} - x \times E_{A} - y \times E_{B}})/(x+y)
\end{equation}

We optimized the metal lattice constants with the computational set up specified in the main manuscript. With this set up, we obtained the following lattice parameters, which we used to set up nanoparticles and slabs. 

\begin{table}[H]
    \centering
    \begin{tabular}{c|c} 
    \hline
    System  & $\mathrm{a \ (\mbox{\AA})}$ \\
    \hline
    Ag & 4.22 \\ 
    Au & 4.20 \\ 
    Cu & 3.66 \\ 
    Pd & 3.99 \\ 
    Pt & 3.99 \\ 
    \hline
    \end{tabular}
\caption{Lattice constants as calculated with our computational set up, see computational details.\label{lattice_constants}}
\end{table}

\section{Machine learning details}\label{ML}
We deployed various machine learning (\textbf{ML}) models, including ordinary linear regression (\textbf{ORL}), Gaussian process regression\cite{rasmussen2006cki} (\textbf{GPR}), neural networks\cite{chollet2015} (\textbf{NN}), random forests (\textbf{RF}), and extreme gradient boost\cite{chen2016xgboost} (\textbf{XGB}) decision trees. We used the scikit-learn packages for ORL/GPR/RF and keras\cite{chollet2015}/ternsorflow for NN, as well as the XGBoost package for XGB. No regularization was used for the ORL. The Gaussian process regression was set up using a linear and a radial basis kernel of length scales between $10^{-3}-10^{3}$. We applied hyper-parameter optimization using the randomized cross-validation (\textbf{cv}) scheme to optimize the XGB and the random forest. Details on the neural network, including the architecture and learning curve are shown in the supporting information, section 3.3. We applied early stopping with a patience of 50 epochs in order to prevent overfitting. 

A genetic algorithm (\textbf{GA}) for feature selection was implemented, the module is available on Github [https://github.com/schlexer/CatLearn]. The algorithm uses a selection of $\mathrm{p}$ features (=genes) from the total feature pool ($\mathrm{p_{tot}}$=28 features) to form a chromosome. A random population of n chromosomes is initialized by randomly selecting chromosomes from the pool of all unique chromosomes; the number of unique chromosomes being given by $\mathrm{\frac{p_{tot}!}{p!(p_{tot}-p)!}}$. For each chromosome, it's fitness (mean of 4f-cv $\mathrm{R^2}$ on the training set) is calculated. The chromosomes are ranked according to their fitness of which m parent chromosomes are selected from the top to create m-1 offspring chromosomes via combination of 2 parent chromosomes, respectively. As the gene order in the chromosome doesn't matter, the parent chromosomes are shuffled before crossover. Then, the first half (or the integer rounding up) of parent 1 is combined with the second half (or the integer rounding down) of parent 2. If any, duplicated genes in the offspring are replaced by a random choice of genes that aren't already in the chromosome (random mutation). Then the fitness of the offspring is calculated and the least well performing m-1 members of the population are replaced by the offspring. The population is ranked again, and the procedure is repeated for g generations. 

Especially for small population sizes $n\rightarrow\mathrm{p_{tot}}$, the occurrence probability of weak genes can drop to 0 and therefore the gene can become absent in the evolution process. In order to prevent this, we monitor the homogeneity index, defined as $\mathrm{1-p_{pop}/p_{tot}}$, where $\mathrm{p_{pop}}$ is the number of unique features in the population. Furthermore, at each generation, after the crossover, a random percentage of chromosomes between 0-20\% are chosen for random mutation. Here, one random gene position is selected and replaced by a random choice of the residual $\mathrm{p_{tot}-p}$ genes which were absent in the original chromosome.

\section{Featurization} \label{features}
\begin{figure}[H]
 \centering
 \includegraphics[width=0.8\linewidth]{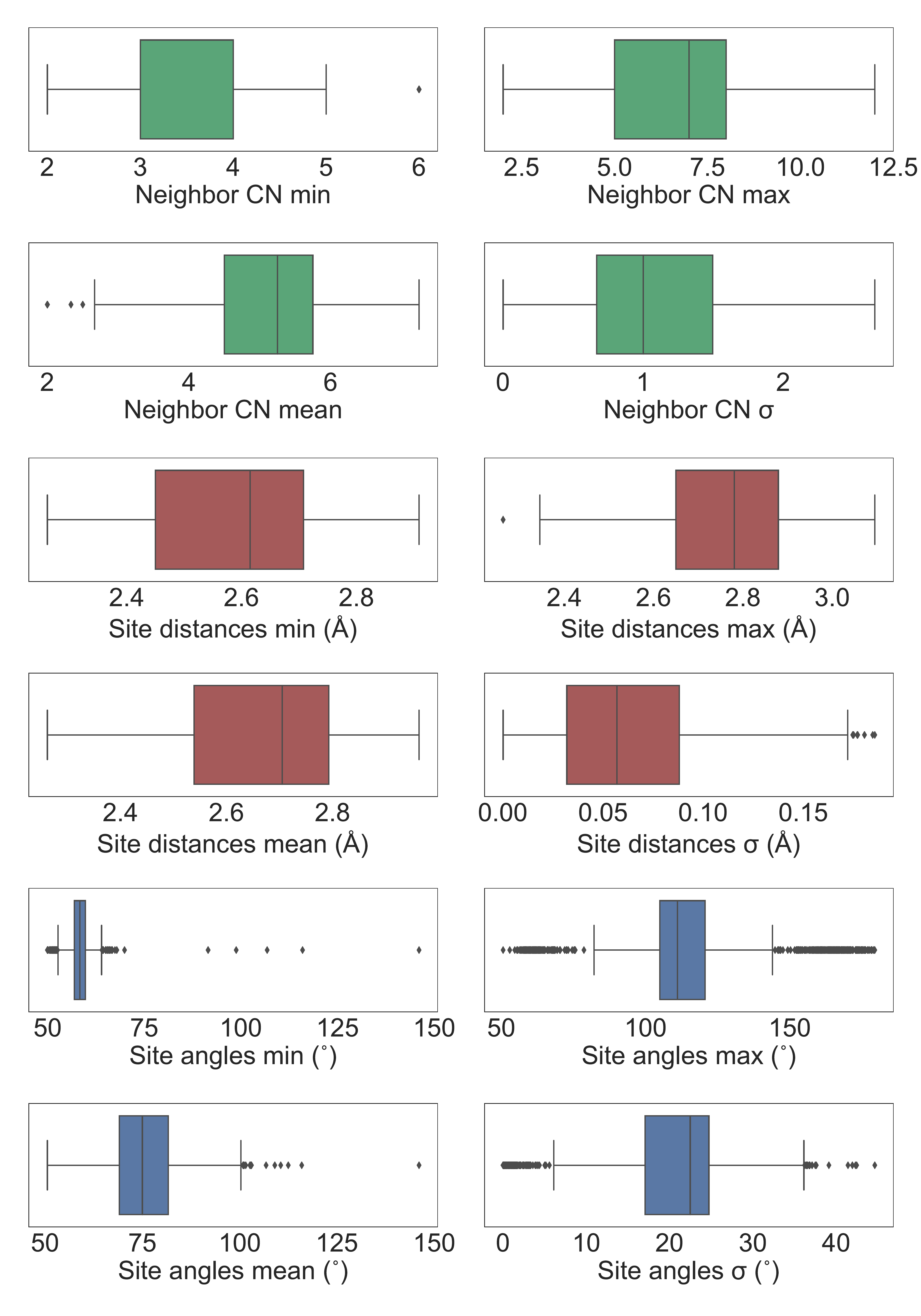}
 \caption{Distributions of site features. Green: Site neighbor coordination number. Red: Site distances. Blue: Site angles. }
 \label{SI_feature_distributions}
\end{figure}

The coordination numbers of atomic sites and the coordination number of their neighbors, as well as relevant statistics thereof where automatically gathered. A neighbor is defined as an atom in a vicinity of 3 $\mathrm{\mbox{\AA}}$. Note that this may lead to more neighbors for smaller atoms. We assume that this discrepancy does not affect the performance of our models, however we abstained from investigating the effect of different neighbor definitions, as this would exceed the scope of this study. We computed not only the coordination number of the site itself, but also that of its neighbors. Chemical intuition suggests that if the neighbors have many neighbors themselves, they would be forming less strong bonds to the site of interest, although of course there can be odd-even effects related to the valence of each respective neighbor, Figure SI-\ref{SI_feature_distributions}.

Site distances are distances between the atomic position of the site and its neighbors, respectively. As each atomic site has several neighbors, statistics like the mean and standard deviation of distances represent the nature of the structure in terms of order/disorder (amorphicity in broader terms) and bond strength. 

Another interesting structural metric is the angle between 3 atoms. Site angles are defined so that the site constitutes the central atom. That is, a site angle is the smallest possible angles measures between $\mathrm{N_{i}-S-N{j}}$, where $\mathrm{N_{i,j}}$ denote the neighbors with $\mathrm{i\neq j}$ and $\mathrm{S}$ denotes the site. 

\begin{figure}[H]
 \centering
 \includegraphics[width=1.00\linewidth]{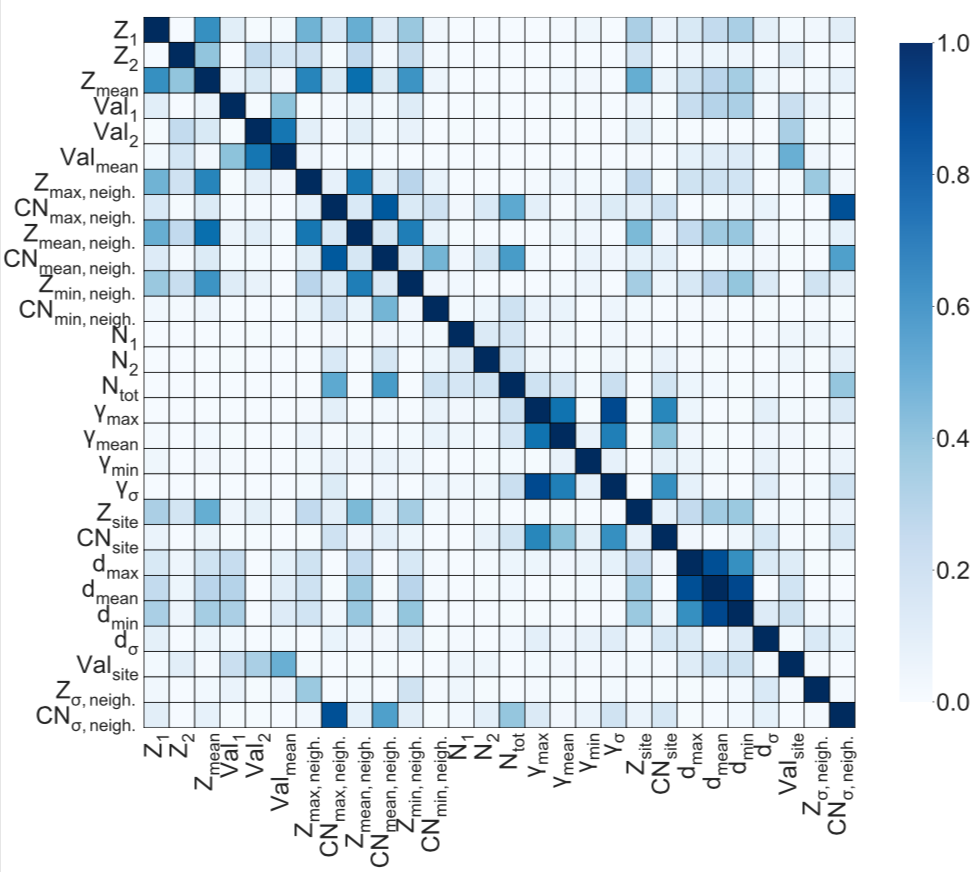}
 \caption{Pearson correlation matrix of features. In bimetallic clusters, element 1 is denoted by the number \textbf{1} and element 2 by the number \textbf{2}. Z is the atomic number, Val is the valence electrons in the elementary state, CN is the coordination number and $\gamma$ is the angle between the atomic site and two neighboring atoms, and d denote inter-atomic distances.}
 \label{SI_correlation}
\end{figure}

As many of the features are derived from the same set of values, such as distance mean, max, standard deviation etc. are measures of the same set of numbers, we can assume some linear correlation between the features. In order to obtain a rough idea on feature correlations, we computed the Pearson feature correlation matrix, shown in Figure \ref{SI_correlation}.

\newpage
\newpage
\section{Model training and feature selection} 
\subsection{Training set size convergence}
\begin{figure}[H]
 \centering
 \includegraphics[width=\linewidth]{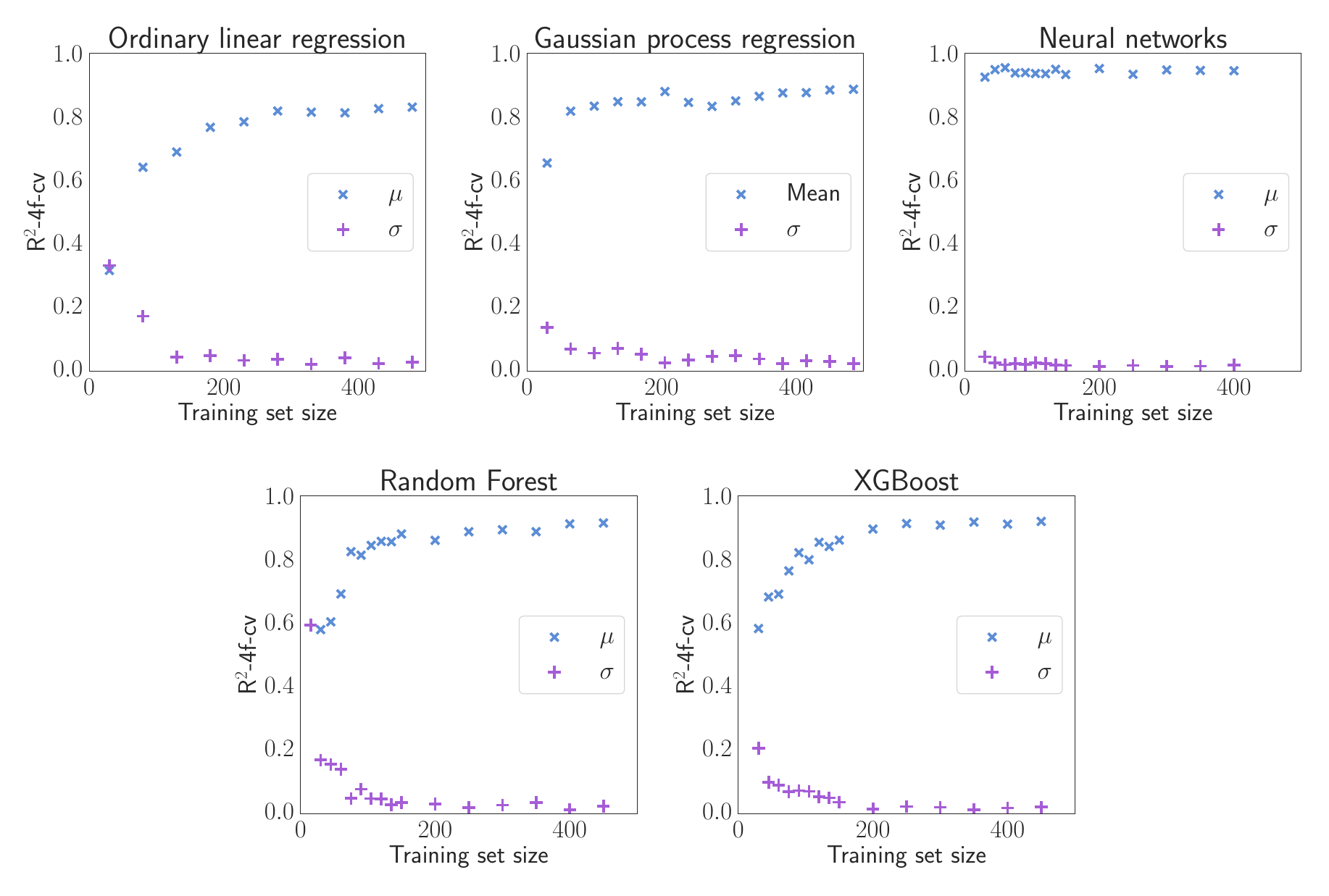}
 \caption{Convergence of mean $\mathrm{R^{2}}$ on 4-fold validation set with training set size.}
 \label{trainin_set}
\end{figure}

We investigated the convergence of our performance metrics which is the mean $\mathrm{R^{2}}$ in 4-fold cross-validation with the training set size, including all features. Interestingly, the data-greedy neural network algorithm already performs well with a smaller training set size. All algorithms reach a plateau at a training set size of 300-400 data points, Figure \ref{trainin_set}.

\subsection{Ordinary linear regression}
\begin{figure}[H]
 \centering
 \includegraphics[width=\linewidth]{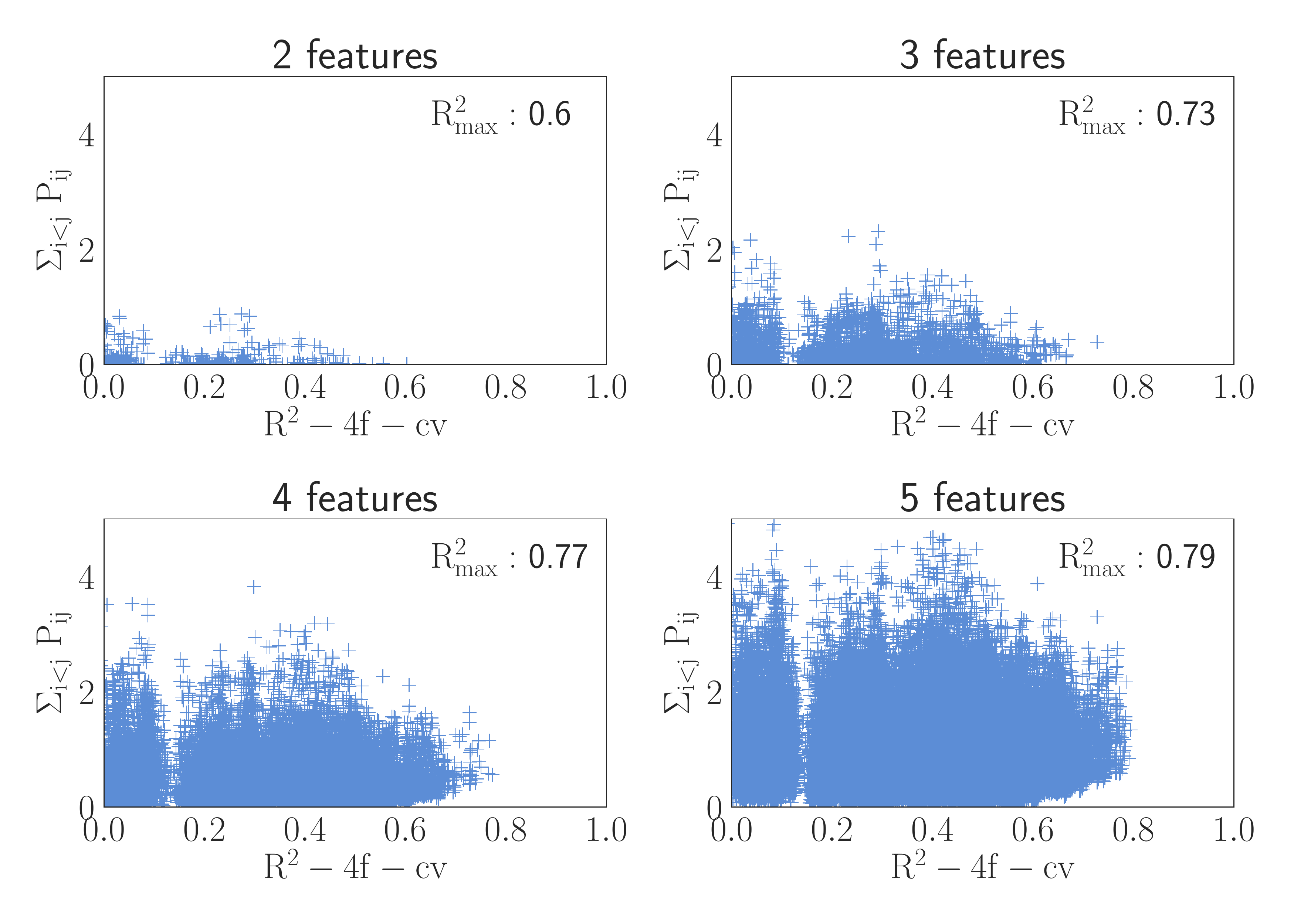}
 \caption{sum of pair-wise Pearson feature correlations, $\mathrm{\Sigma_{i < j }\ P_{ij} < 1}$ and performance metrics mean $\mathrm{R^{2}}$ on 4-fold cross validation set for linear models using 2, 3, 4 and 5 unique features respectively.}
 \label{orl_feature_dim}
\end{figure}

The ordinary linear regression model using all features reaches a $\mathrm{R^{2}}$ value on the test set of 0.81. As some features show collinearity (see Figure \ref{SI_correlation}), we tested, if we could achieve similar performance using less features. To quantify the degree of linear correlations in models, we use the sum of pair-wise Pearson feature correlations, $\mathrm{\Sigma_{i < j }\ P_{ij} < 1}$.  Indeed, using 5 features, the $\mathrm{R^{2}}$ of 0.79 is close to that using all features (0.81), and a few of the solutions showing an $\mathrm{R^{2}}$ close to 0.8 even have small feature correlation $\mathrm{\Sigma_{i < j }\ P_{ij} < 1}$, Figure \ref{orl_feature_dim}. However, for models with 5 features, the majority of the models show significant feature correlation $\mathrm{\Sigma_{i < j }\ P_{ij} > 1}$. We therefore conclude that a feature space of 5 is sufficient to for a subsequent analysis of the importance of individual features.

\subsection{Neural network}
We designed he architecture of the neural network by manual trial and error, comparing cross-validation performances on the training set only. The architecture consists of 6 fully connected layers and one Dropout layer as the second layer. We are aware that there are more sophisticated hyper-parameter optimizations, but as this network performed well, we abstained from further systematic investigations of this matter. We closely monitored the the learning curve of the NN to avoid over-fitting. Using early stopping after 150 epochs* prevents the NN from overfitting to the training data. *During an epoch, the neural net trains on each training point (or batches of training points). After having gone over all training points, the neural net updates it's parameters. Then, it can train again on the training set with the new parameters, which would correspond to the second epoch.

\section{Model performance and feature importance} \label{Models}
\subsection{Clusters}
\begin{figure}[H]
 \centering
 \includegraphics[width=0.99\linewidth]{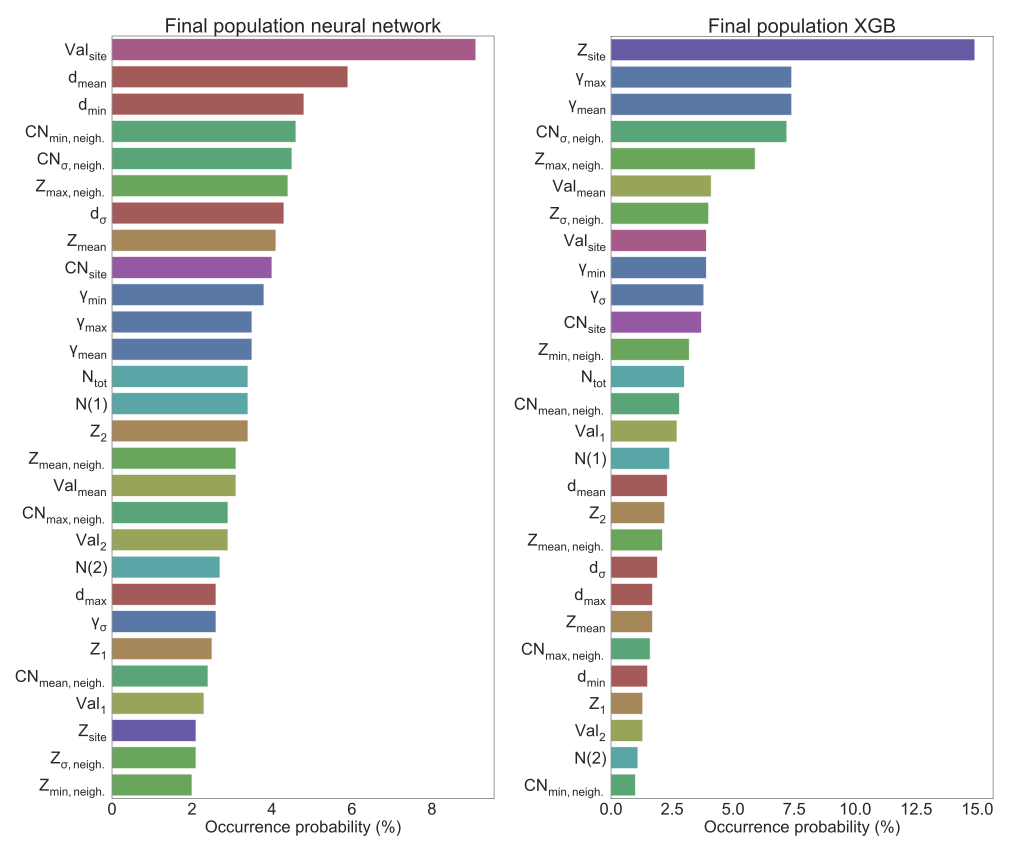}
 \caption{Feature distributions of populations (size 200) after 50 generations of evolution for the neural network (offspring size 10) and the XGB (offspring size 20). }
 \label{gaf_feature_selection}
\end{figure}

\subsection{Nanoparticles}
\begin{figure}[H]
 \centering
 \includegraphics[width=0.9\linewidth]{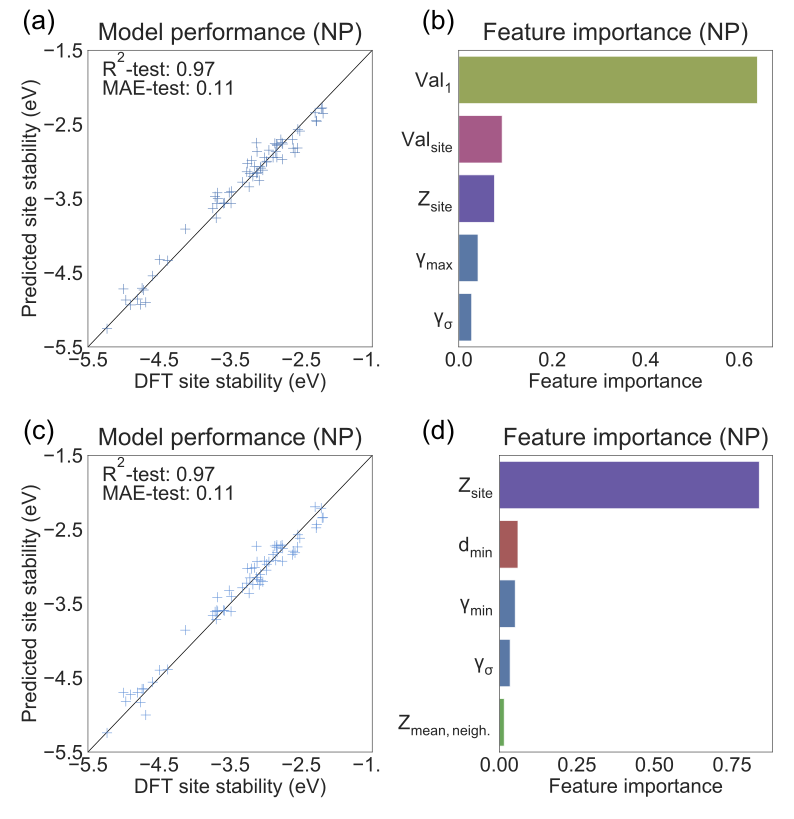}
 \caption{Model performance and feature importance of the extreme gradient boost regressor, training and predicting on the nano-particle data. (a) Parity plot of hyper-optimized XGB using all features. MAE given in eV. (b) Top 5 features of model from (a). (c) Parity plot of hyper-optimized XGB using only 5 features (best model from genetic algorithm evolution). (d) Feature importances of those 5 features used by the best model from (c).}
 \label{np_xgb}
\end{figure}

\subsection{Surfaces}
\begin{figure}[H]
 \centering
 \includegraphics[width=\linewidth]{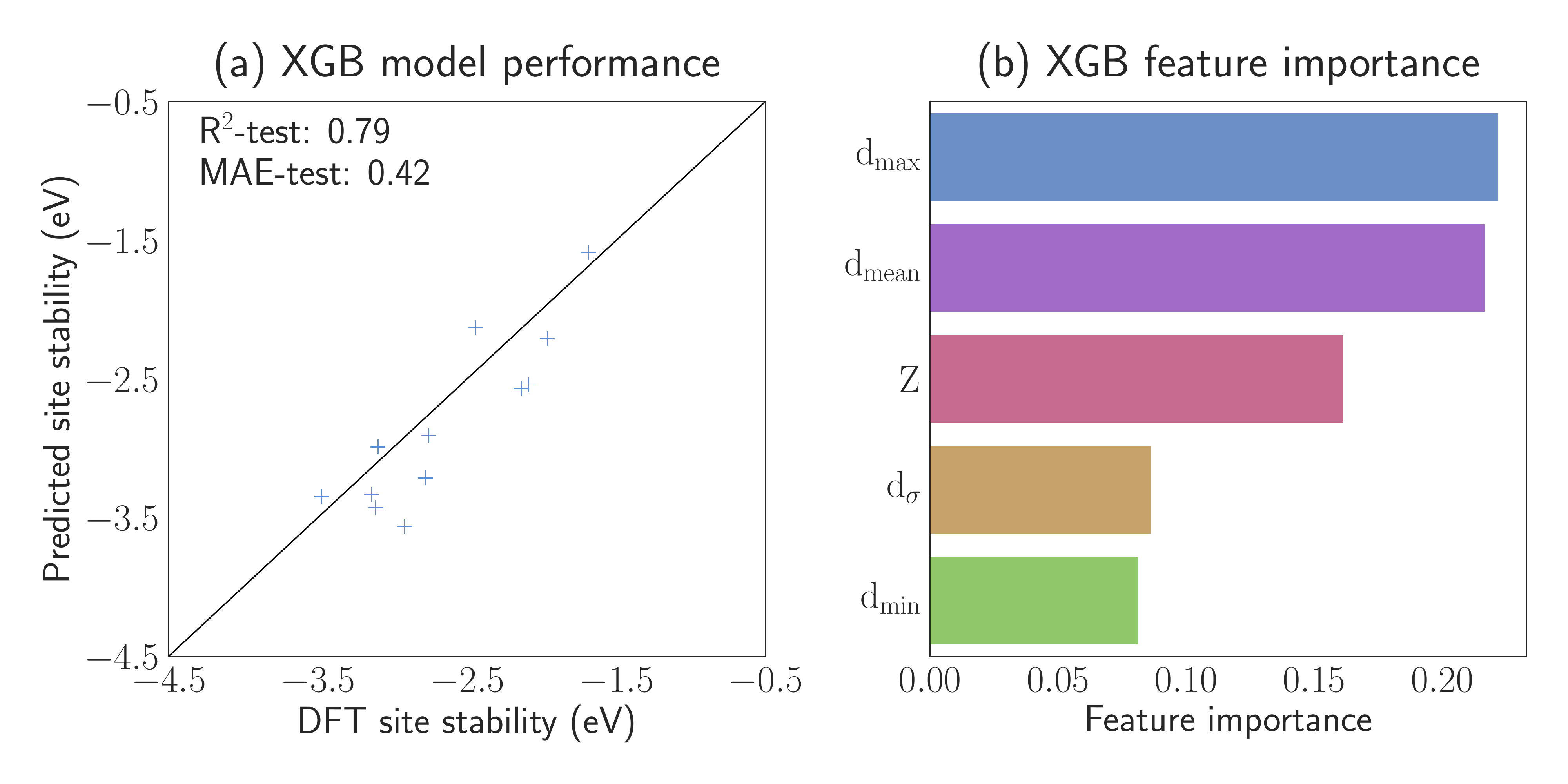}
 \caption{Model performances for surface site stabilities on full feature data set. MAE given in eV. Important features are mainly distance-based.}
 \label{slabs}
\end{figure}

\bibliography{supporting_information.bib}